\documentstyle[12pt,epsf,epsfig]{article}

\newcommand{\p}{\bot}
\newcommand{\dd}{\partial}
\newcommand{\de}{\delta}
\newcommand{\om}{\omega}
\newcommand{\e}{\varepsilon}
\newcommand{\f}{\varphi}
\newcommand{\ls}{\left(}
\newcommand{\rs}{\right)}
\newcommand{\g}{\gamma}
\newcommand{\m}{\mu}
\newcommand{\n}{\nu}
\newcommand{\ra}{\rangle}
\newcommand{\te}{\theta}
\newcommand{\si}{{\rm sign}}
\newcommand{\eel}{e}

\newcommand{\disn}[2]{$$\displaylines{\refstepcounter{equation}%
            \label{#1}\hskip 1em minus 1em #2\hfilneg}$$}
\newcommand{\nom}{\hfil\hskip 1em minus 1em (\theequation)}
\newcommand{\no}{\hfil \hskip 1em minus 1em\phantom{(\theequation)}%
            \hfilneg\cr\hfilneg\hskip 1em minus 1em\hfil}

\makeatletter

\newcount\dobav
\newcount\otlog
\newcount\vrem
\def\@citex[#1]#2{\if@filesw\immediate\write\@auxout{\string\citation{#2}}\fi
  \let\@citea\@empty
  \dobav=-1
  \otlog=-1
  \@cite{\@for\@citeb:=#2\do
    {\def\@tempa##1##2\@nil{\edef\@citeb{\if##1\space##2\else##1##2\fi}}%
     \expandafter\@tempa\@citeb\@nil
     \@ifundefined{b@\@citeb}{\@warning%
       {Citation `\@citeb' on page \thepage \space undefined}%
       \vrem=-1}{\vrem=\csname b@\@citeb\endcsname}
\advance\vrem by -1 \ifnum \vrem=\dobav
 \otlog=\vrem
 \advance\otlog by 1
\else
 \ifnum \vrem=\otlog
  \advance\otlog by 1
 \else
  \ifnum \otlog>0
   \advance\dobav by 1
   \ifnum \otlog=\dobav
    \hbox{,\penalty\@m\ \the\otlog}%
   \else
    \hbox{--\the\otlog}%
   \fi
   \otlog=-1
  \fi
  \dobav=\vrem
  \advance\dobav by 1
  \@citea\def\@citea{,\penalty\@m\ }%
  \ifnum \dobav=-1
   {\reset@font\bf ?}%
  \else
   \hbox{\the\dobav}%
  \fi
 \fi
\fi
}%
\ifnum \otlog>0
 \advance\dobav by 1
 \ifnum \otlog=\dobav
  \hbox{,\penalty\@m\ \the\otlog}%
 \else
  \hbox{--\the\otlog}%
 \fi
\fi }{#1}}

\long\def\@makecaption#1#2{%
   \vskip 10\p@
   \setbox\@tempboxa\hbox{#1. #2}%
   \ifdim \wd\@tempboxa >\hsize
       #1. #2\par
     \else
       \hbox to\hsize{\hfil\box\@tempboxa\hfil}%
   \fi}

\renewcommand{\section}{\@startsection{section}{1}{0pt}%
          {3.5ex plus 1ex minus .2ex}{2.3ex plus .2ex}{\noindent\hfil\bf}}

\makeatother

\begin{document}

\title{
Calculation of the Mass Spectrum of QED-2\\
in Light-Front Coordinates\\}

\author{S.~A.~Paston\thanks{E-mail: paston@pobox.spbu.ru},
E.~V.~Prokhvatilov\thanks{E-mail: Evgeni.Prokhvat@pobox.spbu.ru},
V.~A.~Franke\thanks{E-mail: franke@pobox.spbu.ru}\\
{\it St.-Petersburg State University, Russia}}

\date{\vskip 15mm}

\maketitle

\begin{abstract}
With the aim of a further investigation of the nonperturbative
Hamiltonian approach in gauge field theories, the mass spectrum
of QED-2 is calculated numerically by using the corrected
Hamiltonian that was constructed previously for this theory on
the light front. The calculations are performed for a wide range
of the ratio of the fermion mass to the fermion charge at all
values of the parameter $\hat\te$ related to the vacuum angle $\te$.
The results obtained in this way are compared with the results of
known numerical calculations on a lattice in Lorentz coordinates.
A method is proposed for extrapolating the values obtained within
the infrared-regularized theory to the limit where the
regularization is removed. The resulting spectrum agrees well
with the known results in the case of $\te=0$; in the case of $\te=\pi$,
there is agreement at small values of the fermion mass (below
the phase-transition point).
\end{abstract}

\newpage
\section{Introduction}
The Hamiltonian approach to quantum-field theory in light-front
coordinates $x^{\pm}=(x^0 \pm x^3)/\sqrt{2},\; x^{\p}=
(x^1,x^2)$, where $x^+$ plays the
role of time \cite{1}, is one of the nonperturbative methods for
solving the problem of strong interaction \cite{3,4}. Within this
approach, the quantization is performed in the $x^+=0$ plane, the
generator $P_+$ of a shift along the $x^+$ axis playing the role of the
Hamiltonian. The generator $P_-$ of a shift along the $x^-$ axis does
not displace the quantization surface; therefore, it is a
kinematical generator (according to Dirac's terminology) in
contract to the dynamical generator $P_+$. As a result, the momentum
operator $P_-$ appears to be quadratic in fields and does not depend
on interaction. At the same time, the operator $P_-$ is nonnegative
and has zero eigenvalue only on the physical vacuum. This results
in that the field Fourier modes corresponding to positive and
negative values of $p_-$ play the role of, respectively, creation
and annihilation operators over the physical vacuum and can be
used to construct Fock space. Thus, we see that, in light-front
coordinates, the physical vacuum formally coincides with the
mathematical vacuum.

The spectrum of bound states in the theory can be sought by
solving the Schroedinger equation
 \disn{1}{
P_+|\Psi\rangle =p_+|\Psi\rangle
 \nom}
in the subspace specified by fixed $p_-$ and $p_{\p}$ and by employing the
expression $m^2= 2p_+p_- -p_{\p}^2$ for the mass. This search for bound
states can be performed beyond perturbation theory -- for example,
with the aid of so-called method of discrete light-cone
quantization \cite{3,5}.

However, the light-front Hamiltonian formalism involves a
specific divergence at $p_-=0$ \cite{3,4}, and it must be regularized.
The introduction of a cutoff $|p_-|\ge\e >0$, which violates Lorentz
and gauge invariance, is one of the methods for its
regularization. A cutoff $|x^-|\le L$ that involves imposing
(anti)periodic boundary conditions in $x^-$ (discrete light-cone
quantization method, which respects gauge invariance) is yet
another possible regularization. In this case, the lightlike
momentum $p_-$ becomes discrete ($p_-=p_n=\pi n/L$, where $n$ is an
integer), the field zero mode corresponding to $n=0$ being
separated explicitly. In principle, the canonical formalism makes
it possible to express this zero mode in terms of other modes by
solving the constraint equation, but this is difficult as a rule
\cite{fnp1,fnp2}.

The regularization of the above divergence usually renders a
theory in light-front coordinates nonequivalent to its
conventional formulation in Lorentz coordinates \cite{ann,14,naus}. This can
be revealed even in the lowest orders of perturbation theory
\cite{17}. As a result, there arises the problem of correcting the
canonical light-front Hamiltonian (which is the result of a
"naive" canonical quantization in light-front coordinates) -- that
is, the problem of seeking counterterms to it that compensate for
the above distinctions between the Hamiltonians. If this problem
can be solved for a specific theory in all orders of perturbation
theory, the resulting corrected light-front Hamiltonian can then
be used to perform nonperturbative calculations.

The aforementioned formal coincidence of the physical and the
mathematical vacuum becomes
rigorous only after the introduction of regularization, upon
which the vicinity of the point $p_-=0$ is eliminated -- for example,
after the introduction of the cutoff $|p_-|\ge\e >0$. If the corrected
light-front Hamiltonian can be constructed for a theory
regularized in this way, then vacuum effects inherent in the
original theory in Lorentz coordinates must be taken into account
with the aid of additional terms of this Hamiltonian.

The problem of constructing the corrected canonical light-front
Hamiltonian was successfully solved both for nongauge field
theories of the Yukawa model type \cite{20} and for QCD in the gauge
$A_-=0$ \cite{21}. In the last case, however, the corrected light-front
Hamiltonian was constructed only for specific ultraviolet and
infrared regularizations violating gauge invariance. As a result,
it appears that the corrected Hamiltonian involves a large number
of indeterminate coefficients; only for some, a priori unknown,
dependence of these coefficients on the regularization parameter
does it reproduce, in the limit where the regularization is
removed, the results of the Lorentz-covariant theory in all
orders of perturbation theory. The practical calculations with
the resulting Hamiltonian are very cumbersome because of the
presence of unknown coefficients and because of a complicated
structure of regularization (the regularized Hamiltonian involves
a large number of additional fields).

In view of these circumstances, it is desirable to seek
alternative methods for constructing the correct light-front
Hamiltonian for gauge theories. In this connection, it is of
interest to study the simplest models that admit a
nonperturbative approach -- in particular, those where one can
study the behavior of infinite series of perturbation theory in
all orders. Two-dimensional QED (QED-2) featuring a nonzero
fermion mass (it is also known as the massive Schwinger model) is
one of such models. In recent years, this two-dimensional model
has attracted attention as an object of application of new
methods for studying QCD, since it possesses many properties
similar to those of QCD: confinement, chiral-symmetry breaking,
and a topological $\te$ vacuum (see \cite{ham2} and references therein, as
well as \cite{ham1,adam1,adam2}). Information obtained in analyzing QED-2 can
also be used in developing new methods that take into account
nonperturbative vacuum effects and which are appropriate for
constructing the light-front Hamiltonian for four-dimensional
gauge theories. It should be noted that attempts at extracting
information about four-dimensional gauge theories on the light
front from an analysis of QED-2 were undertaken earlier \cite{mac1}.

For QED-2, there exists the possibility of going over to an
equivalent scalar theory \cite{24} (belonging to the type of the
sine-Gordon model). This can be done by means of the bosonization
procedure -- that is, by going over from the fermion variables to
boson ones \cite{naus,28}. Upon this transition, the mass term of the
fermion field in the QED-2 Hamiltonian becomes the interaction
term for a scalar field, while the fermion mass $M$ becomes the
interaction constant in the boson theory. In the boson theory,
the fact that the quantum vacuum in QED-2 has a nontrivial
character associated with instantons ($\te$ vacuum) \cite{24,adam1} is taken
explicitly into account with the aid of the parameter $\te$ in the
interaction term. At $M=0$, QED-2 reduces to the Schwinger model,
while the equivalent boson theory appears to be free.

Perturbation theory for a boson theory (perturbation theory in
the fermion mass) is usually referred to as chiral perturbation
theory. For this kind of perturbation theory, ultraviolet
finiteness was proven in \cite{tmf03,shw2}. By analyzing perturbation
theory in all orders of $M$, one can construct a corrected
light-front Hamiltonian in terms of bosons and, after this,
return to the fermion variables \cite{shw2,tmf02}. It should be noted that
boson perturbation theory differs radically from perturbation
theory in the coupling constant of the original theory involving
fermions (in QED-2, the latter perturbation theory does not exist
at all because of infrared divergences, and this was the reason
for introducing bosonization). Therefore, the resulting
light-front Hamiltonian can take into account nonperturbative (in
the conventional coupling constant) effects. But at the same
time, it can fail to describe effects that are nonperturbative in
the fermion mass -- for example, phase transitions.

It is well known that, at least at the vacuum-angle value of $\te=\pi$,
there is a phase transition in QED-2 at some value of the
fermion mass $M$ (see, for example, \cite{ham2}). It should be expected
that, in the presence of a phase transition, which is accompanied
by the appearance of nonzero vacuum expectation values of some
operators, the correct light-front Hamiltonian must have
different form for different phases, since the light-front vacuum
itself is always trivial. Therefore, the results of calculations
performed with a specific Hamiltonian must be valid only within
one phase. The calculations performed in the present study
corroborate these considerations and make it possible to
determine the presence and an approximated position of the
phase-transition point. A similar phenomenon was discovered
previously in the simple two-dimensional $\lambda\f^4$ scalar-field model
(see \cite{naus,pred}, where an approximate method was proposed for going
over to the Hamiltonian describing a different phase).

In this study, we perform a numerical nonperturbative calculation
of the mass spectrum of the corrected light-front QED-2
Hamiltonian constructed in \cite{shw2,tmf02}. The results obtained in this
way are compared with the results of numerical calculations on a
lattice in Lorentz coordinates from \cite{ham1,ham2}.

\section{Corrected light-front Hamiltonian for QED-2}
The QED-2 Lagrangian density in Lorentz coordinates has the form
 \disn{2}{
L=-\frac{1}{4}F_{\m\n}F^{\m\n}+\bar\Psi(i\g^\m D_\m-M)\Psi,
\nom}
where $F_{\mu\nu}= \dd_{\mu} A_{\nu}- \dd_{\nu }A_{\mu}$,
$D_{\mu}=\dd_{\mu} - i\eel A_{\mu} $,  $A_{\mu}(x)$
is an Abelian gauge field, $\Psi$ and $\bar\Psi =\Psi^+\g^0$
are two-component fermion fields of mass $M$, $\eel$ is the coupling
constant, and the matrices $\g^\m$ are chosen in the form
 \disn{2.2.1}{
\g^0=\ls
\begin{array}{cc}
0 & -i\\
i & 0
\end{array}
\rs\! , \quad \g^1=\ls
\begin{array}{cc}
0 & i\\
i & 0
\end{array}
\rs\! .
\nom}
In terms of Lorentz coordinates, the Lagrangian density for the
boson theory equivalent to QED-2 can be written in the form \cite{shw2}
 \disn{3.0}{
 L=\frac{1}{8\pi}\ls\dd_\m\f\dd^\m\f-m^2\f^2\rs+
 \frac{\g}{2}e^{i\te}:e^{i\f}:+\frac{\g}{2}e^{-i\te}:e^{-i\f}:,\qquad
 \g=\frac{Mme^C}{2\pi},\quad
 m=\frac{\eel}{\sqrt\pi},
 \nom}
where $C=0.577216$ is the Euler constant, $\te$ is a quantity
that parametrizes the $\te$ vacuum of the fermion formulation of
the theory, and the normal-ordering symbol means that diagrams
with connected lines are excluded in perturbation theory in $\g$
(this corresponds to the usual meaning of the normal-ordering
symbol in the Hamiltonian) -- it is equivalent to perturbation
theory in $M$.

In \cite{shw2,tmf02}, the light-front Hamiltonian generating a theory that
describes a one-component fermion field $\psi$ and which, in the limit
where the regularization is removed, is equivalent in all orders
in $\g$ to the Lorentz-covariant theory specified by the Lagrangian
density (\ref{3.0}) was found by using the method described in the
Introduction. The theory defined by this Hamiltonian was
regularized by the discrete-light-cone-quantization method
mentioned in the Introduction: the cutoff $|x^-|\le L$ and the
antiperiodic boundary conditions in $x^-$ were introduced for the
field $\psi$. The resulting corrected light-front Hamiltonian has the
form
 \disn{3.1}{
 H=\int\limits_{-L}^Ldx^-\biggl(\frac{\eel^2}{2}\ls \dd_-^{-1}
 [\psi^+\psi]\rs^2-\frac{\eel Me^C}{4\pi^{3/2}}
 \ls
 e^{-i\hat\te(M/\eel,\:\te)}
 \:e^{i\om}d_0^++h.c.\rs-
 \frac{iM^2}{2}\psi^+\dd_-^{-1}\psi\biggr),
 \nom}
where $[\dots]$ denote the omission of the zero mode in $x^-$. The
field $\psi$ is expanded in terms of creation and annihilation
operators as
 \disn{3.2}{
 \psi(x)=\frac{1}{\sqrt{2L}}\ls \sum_{n\ge 1}b_n
 e^{-i\frac{\pi}{L}(n-\frac{1}{2})x^-}+ \sum_{n\ge 0}d_n^+
 e^{i\frac{\pi}{L}(n+\frac{1}{2})x^-}\rs\! ,
 \nom}
 \disn{3.21}{
 \{b_n,b_{n'}^+\}=\{d_n,d_{n'}^+\}=\de_{nn'},\quad
 b_n|0\ra=d_n|0\ra=0.
 \nom}
The operator $\om$ is the quantity canonically conjugate
to the charge operator $Q$
 \disn{3.3}{
 Q=\sum_{n\ge 1}b_n^+b_n-\sum_{n\ge 0}d_n^+d_n,
 \nom}
which specifies the physical subspace of states, $|{\rm phys}\ra$:
 \disn{3.4}{
 Q |{\rm phys}\ra=0.
 \nom}
The operator $\om$ possesses the properties that completely define
it \cite{naus,28},
 \disn{3.5}{
e^{i\om}|0\ra=b^+_1 |0\ra,\quad e^{-i\om}|0\ra=d^+_0 |0\ra
 \nom}
and
 \disn{3.6}{
e^{i\om}\psi(x)e^{-i\om} =e^{i\frac{\pi}{L}x^-}\psi(x),
 \nom}
whence it follows that
 \disn{3.7}{
 e^{i\om}b_n e^{-i\om}=b_{n+1},\quad
 e^{i\om} d_n^+ e^{-i\om}=d_{n-1}^+,\quad  n\ge 1,\quad\qquad
 e^{i\om} d_0^+ e^{-i\om} =b_1.
 \nom}

The parameter $\hat\te$ appearing in the Hamiltonian in (\ref{3.1}) is a
function of the ratio $M/\eel$ and the vacuum angle $\te$. This function
is defined as a perturbation-theory series in $M$; therefore, its
an explicit form remains unknown. Details concerning the
appearance of the parameter $\hat\te$ in the Hamiltonian are considered
in the Appendix. Among other things, it is established there
that, in the first-order in $M$, we have $\hat\te=\te$ and that, at any
value of $M$, the parameter $\hat\te$ is an odd function
of $\te$ and takes the
value of $\hat\te=\pi$ at $\te=\pi$. In particular, it follows from the
oddness of the function $\hat\te(\te)$ that $\hat\te=0$ at $\te=0$.
It should be
noted that the parameter $\hat\te$ can be related to the values of the
vacuum condensates in the Lorentz-covariant theory \cite{shw2,tmf02}.

In calculating the mass spectrum of bound states, the quantity $\hat\te$
is an independent parameter of the theory, along with $M$ and $\eel$;
the relation between $\hat\te$ and $\te$ for $\te\ne 0,\pi$
can in principle be
found by comparing the results obtained by calculating the mass
spectrum of the theory in Lorentz coordinates and the theory on
the light front. We note that expression (\ref{3.1}) for the corrected
light-front Hamiltonian and the expression obtained upon the
naive canonical quantization of the original fermion theory (\ref{2})
in the light-front coordinates differ only by the addition of the
second term, which is linear in the field operators and which
depends on $\hat\te$ and, hence, on the vacuum angle $\te$. Thus, we see that
the naive canonical quantization does not take into account
vacuum effects.

The lightlike-momentum operator $P_-$ has the form
 \disn{3.8}{
 P_- =\sum_{n\ge 1}b_n^+b_n\frac{\pi}{L}\ls n-\frac{1}{2}\rs+
 \sum_{n\ge 0}d_n^+d_n\frac{\pi}{L}\ls n+\frac{1}{2}\rs.
 \nom}
This expression is used to calculate the mass spectrum of bound states.

\section{Calculation of the mass spectrum of bound states}
In order to find the mass spectrum of bound states of the theory,
we will seek the eigenvalues $E_i$ of the fermion light-front
Hamiltonian (\ref{3.1}) (the subscript $i$ numbers eigenvalues in the
ascending order),
 \disn{4.1}{
H|\Psi_i\ra=E_i|\Psi_i\ra
\nom}
in the subspace of physical states at a fixed
value of the lightlike momentum (\ref{3.8}). This subspace is specified
by the conditions
 \disn{4.2}{
 Q |\Psi\ra=0, \qquad P_-|\Psi\ra=p_-|\Psi\ra.
 \nom}
In view of the antiperiodic boundary conditions and the first of
the equalities in (\ref{4.2}), the eigenvalue $p_-$ has the form
 \disn{4.3}{
 p_-=\frac{\pi}{L}N,
 \nom}
where $N$ is a nonnegative integer. The bound-state masses $M_i$ are given by
 \disn{4.4}{
 M_i^2=2p_-E_i=\frac{2\pi}{L}NE_i.
 \nom}
If, in expression (\ref{3.1}) for the Hamiltonian, one performs
the change of integration variable $x^-=\frac{L}{\pi}z$ and uses
expansion (\ref{3.2}), the operator $\frac{2\pi}{L}NH$,
which determines the quantities $M_i^2$ does not involve
the regularization parameter $L$ explicitly, but it depends
on $N$. The quantity $L$ affects the mass spectrum only through
relation (\ref{4.3}). Since $p_-$ does not depend on $L$,
one can deduce from (\ref{4.3}) that the limit $N\to\infty$ corresponds
to the limit where the regularization is removed, $L\to\infty$.

Since we use the antiperiodic boundary conditions, the subspace
specified by the conditions in (\ref{4.2})) appears to be finite. This
occurs because there exists a minimum positive value of the
lightlike momentum $p_-=\frac{\pi}{2L}$ and because the creation
operators correspond only to positive values of $p_-$. An arbitrary
state satisfying the conditions in (\ref{4.2}) has the form
 \disn{4.5}{
|\Psi\ra={d_{N-1}^+}^{l_{2N}}\dots{d_{0}^+}^{l_{N+1}}
{b_{1}^+}^{l_{N}}\dots{b_{N}^+}^{l_{1}}|0\ra,
\qquad l_i=0,1;\no
\sum_{i=1}^{2N}\si\ls N-i+\frac{1}{2}\rs l_i=0,\qquad
\sum_{i=1}^{2N}\left| N-i+\frac{1}{2}\right| l_i=N,
\nom}
where sgn is the sign function. The values $l_i$ takes here only
the values of $0$ or $1$ by virtue of the anticommutation relations
(\ref{3.21}).

The finiteness of the subspace that is specified by formulas (\ref{4.5})
and where it is necessary to solve Eq.~(\ref{4.1}) reduces the problem
to finding the eigenvalues of an $N_{{\rm mat}}\times N_{{\rm mat}}$
finite matrix whose
elements are specified by the matrix elements of the operator
$\frac{2\pi}{L}NH$ between the states in (\ref{4.5}).
A precise solution to this
problem can be found numerically. The resulting eigenvalues will
determine the squares $M_i^2$ of bound-state masses. It should be
noted that the dimension of the matrix, $N_{{\rm mat}}$, grows fast as the
parameter $N$ increases -- this is reflected in Table~1.
\begin{table}[ht]
\begin{center}
\begin{tabular}{|c|c|c|c|c|c|c|c|c|c|c|c|c|c|}
\hline
$N$&9&10&12&14&16&18&20&22&24&26&28&29&30\\
\hline
$N_{{\rm mat}}$&31&43&78&136&232&386&628&1003&1576&2437&3719&4566&5605\\
\hline
\end{tabular}
\caption{
Relation between the parameter $N$ and the dimensionality $N_{{\rm mat}}$ of
the space of states.
}
\end{center}
\end{table}
In this
study, the maximum achieved values of $N$ are $N=30$ for the cases
of $\hat\te=\te=0$ and $\hat\te=\te=\pi$ and N = 28 for the remaining cases.

As was mentioned above, the limit $N\to\infty$ corresponds to the
removal of the regularization. For this reason, it is not
sufficient to calculate the mass spectrum $M_i^2$ at the maximum
accessible value of $N$ -- it is necessary to analyze the behavior of
the spectrum as a function of $N$ and to find the way to
extrapolate the calculated values to the region of $N\to\infty$. We
propose the following method of extrapolation. We introduce the
quantity $u=1/N$ and consider the function $M_i^2(u)$. It is necessary
to extrapolate the values of this function to zero. Our
calculations reveal that the bound-state masses are sensitive to
the parity of $N$ -- that is, there can occur a sharp change in $M_i^2$ in
response to the reversal of the parity of $N$. Therefore, it is
reasonable to extrapolate the function
$M_i^2(u)$ to zero by two methods, individually in even
and in odd values of $N$.

In order to extrapolate the function $M_i^2(u)$ to zero, we
approximate the dimensionless ratio $M_i^2(u)/\eel^2$ by polynomials of
various degrees by the least squares method. We denote by $P_i(n)$
the value of a polynomial of degree $n$ at zero. It is obvious that
the maximum degree $n$ that can be used is less by unity than the
number of points at which the approximated function is known. In
the present study, this degree is equal to 10 for the cases of
$\hat\te=\te=0$ and $\hat\te=\te=\pi$ and to 9 in the remaining cases (as the
minimum value of $N$, we adopt $N=9$ in our calculations).

At different values of the ratio $M/\eel$ and the parameter $\hat\te$, there
arise different types of behavior of $P_i(n)$ as a function of $n$. In
some cases, the function $P_i(n)$ tends to a saturation and changes
slowly with increasing $n$ (see Fig.~l$a$).
\begin{figure}[htb]
\begin{center}
\epsfig{file=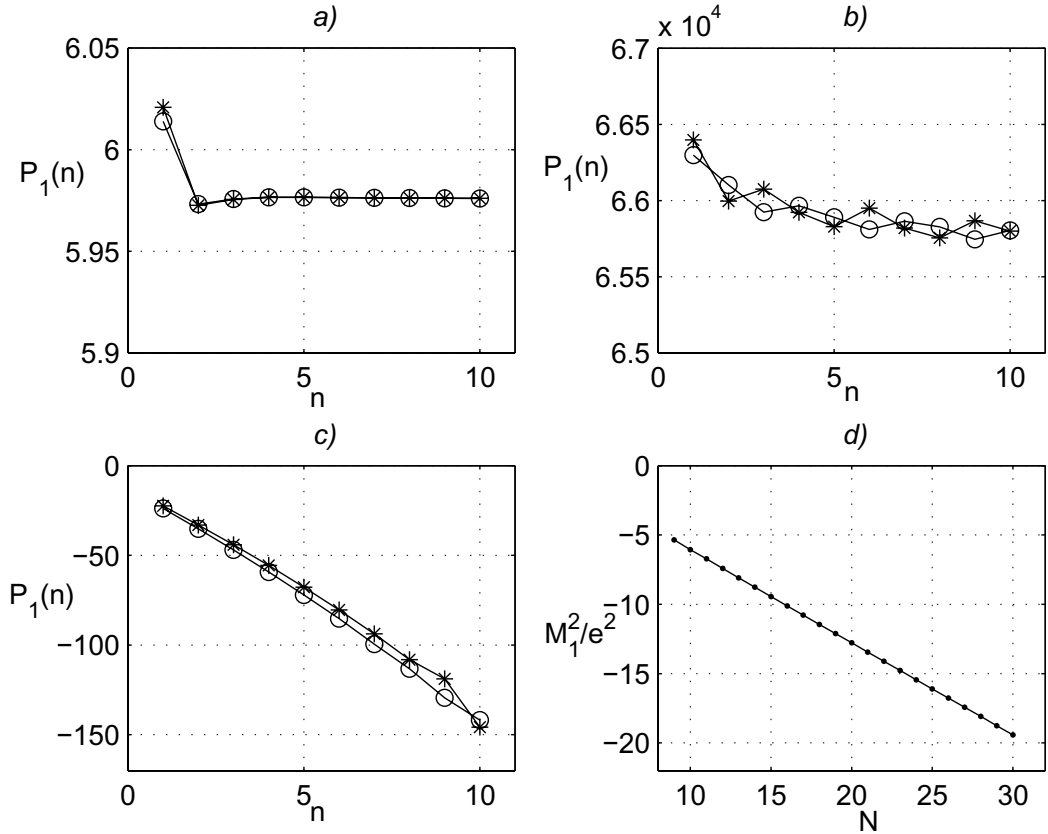,width=14cm}
\caption{
Examples of the dependence of the extrapolated value of $P_1(n)$ on
the degree $n$ of the approximating polynomial ($a$-$c$) and example of
the dependence of the mass of the lowest bound state on N ($d$) for
the following parameter values: ($a$) $\hat\te=\te=0$ and $M/\eel=1$,
($b$) $\hat\te=\te=0$ and $M/\eel=2^7$,
and ($c$,$d$) $\hat\te=\te=\pi$ and $M/\eel=0.5$. The symbols
$\circ$ and $*$ represent the results of the
extrapolation in, respectively, even and odd values of $N$.
}
\end{center}
\end{figure}
In these cases, the value
at which the saturation occurs will be considered as the result
of extrapolating the function $M_i^2(u)/\eel^2$ to zero.

Sometimes, there arise oscillations against the background of a
saturation (see Fig.~1$b$). This occurs if the error with which one
calculates $M_i^2(u)$ becomes sizable. In this case, a function that
involves a noticeable random noise is approximated by a
polynomial of high degree. This usually takes place at large
values of the ratio $M/\eel$, in which case the coefficients of
various terms of the Hamiltonian (\ref{3.1}) differ from one an other
considerably. Conceptually, this situation does not differ from
the preceding one. The value obtained by averaging the
oscillations in the region of saturation will then be treated as
the result of the extrapolation of the function $M_i^2(u)/\eel^2$
to zero.
It is obvious that the value found in this way will have an error
larger than that in the preceding situation.

In the remaining cases, there arises the situation where the
saturation cannot be seen -- the function $P_i(n)$ strongly changes
with increasing $n$ (see Fig.~1$c$). In order to discriminate between
this case and the two preceding versions of behavior of the
function $P_i(n)$, we introduce the measure of its relative
variation in response to a considerable variation in the degree $n$
of the polynomial (in the region of accessible values), for
example, in the form
 \disn{4.6}{
\xi=\sqrt{\frac{\ls P_i(4)-P_i(9)\rs^2}{\ls P_i(4)^2+P_i(9)^2\rs /2}}.
\nom}
This quantity characterizes the error with which the calculated
values of the mass spectrum $M_i^2$ describe its limiting value. We
will assume that, if $\xi<0.1$, the saturation takes place for the
function $P_i(n)$, so that its value obtained at accessible $n$
describes well the limiting value of the bound-state mass. But if
$\xi>0.1$, the calculated values of $P_i(n)$ do not characterize the
behavior of $M_i^2$ in the limit $N\to\infty$. Moreover, the most
frequently occurring form of the dependence $P_i(n)$ suggests its
linearly decreasing character (Fig.~1$c$). The same applies to the
original dependence of the quantity $M_i^2$ on the parameter $N$ (see
Fig.~1$d$). This gives sufficient grounds to assume that, at these
values of the ratio $M/\eel$ and the parameter $\hat\te$, the quantity $M_i^2$
tends to $-\infty$ in the limit $N\to\infty$. The possible reasons behind
this effect are discussed in the next section.

It should be noted that the choice of the value of 0.1 as a
boundary one for the error in $\xi$ and the specific choice of formula
(\ref{4.6}) are arbitrary to a considerable extent, but, unfortunately,
our calculations could not provide a more rigorous way to
discriminate between the situations where the limit of the
quantity $M_i^2$ for $N\to\infty$ exists and where $M_i^2$
tends to $-\infty$ in the same limit.

\section{Results of the calculations}
{\bf 4.1. Case of $\hat\te=\te=0$.}

In the case of $\te=0$, the mass spectrum of the massive
Schwinger model in Lorentz coordinates has received quite an
adequate study (see \cite{ham2,adam2} and references therein). Usually,
one studies the masses $M_1$ and $M_2$ of the first two bound states,
which are referred to as a vector and a scalar state,
respectively. The most accurate results were obtained in \cite{ham1}
with the aid of lattice calculations. Table~2 presents the values
of $(M_1-2M)/\eel$ and $(M_2-2M)/\eel$ (it is precisely these quantities
that were calculated in \cite{ham1}) that were found by the method
proposed here (with the aid of an extrapolation to the limit $N\to\infty$)
and the values of these quantities from \cite{ham1}.
\begin{table}[htb]
\begin{center}
\begin{tabular}{|c|c|c|c|c|}
\hline
$M/\eel$ & \multicolumn{2}{c|}{$(M_1-2M)/\eel$} & \multicolumn{2}{c|}{$(M_2-2M)/\eel$}\\
\cline{2-5}
& our study & \protect\cite{ham1} & our study & \protect\cite{ham1} \\
\hline
$2^{-10}$ & 0.564 &        &  1.13 &      \\
\hline
$2^{-9}$  & 0.564 &        &  1.13 &      \\
\hline
$2^{-8}$  & 0.563 &        &  1.13 &      \\
\hline
$2^{-7}$  & 0.563 &        &  1.14 &      \\
\hline
$2^{-6}$  & 0.561 &        &  1.15 &      \\
\hline
$2^{-5}$  & 0.559 &        &  1.17 &      \\
\hline
$2^{-4}$  & 0.554 &        &  1.20 &      \\
\hline
$2^{-3}$  & 0.545 & 0.543  &  1.23 & 1.22 \\
\hline
$2^{-2}$  & 0.524 & 0.519  &  1.24 & 1.24 \\
\hline
$2^{-1}$  & 0.489 & 0.485  &  1.20 & 1.20 \\
\hline
$2^{0}$   & 0.445 & 0.448  &  1.12 & 1.12 \\
\hline
$2^{1}$   & 0.393 & 0.394  &  0.99 & 1.00 \\
\hline
$2^{2}$   & 0.339 & 0.345  &  0.84 & 0.85 \\
\hline
$2^{3}$   & 0.295 & 0.295  &  0.75 & 0.68 \\
\hline
$2^{4}$   & 0.279 & 0.243  &  0.74 & 0.56 \\
\hline
$2^{5}$   & 0.302 & 0.198  &  0.84 & 0.45 \\
\hline
$2^{6}$   & 0.368 &        &  1.08 &      \\
\hline
$2^{7}$   & 0.497 &        &  1.41 &      \\
\hline
$2^{8}$   & 0.619 &        &       &      \\
\hline
\end{tabular}
\caption{
Masses of the vector ($M_1$) and scalar ($M_2$) bound states according
to the calculations performed in the present study and in
\protect\cite{ham1} at
$\hat\te=\te=0$ and various values of the ratio $M/\eel$.
}
\end{center}
\end{table}
In the case of
$\te=0$, the error $\xi$ does not exceed the threshold of 0.1 at any
values of the ratio $M/\eel$ that were considered here (more
specifically, $\xi<0.01$ for $M_1$ and $\xi<0.03$ for $M_2$) --
that is, the above procedure of extrapolation to the limit $N\to\infty$
provides quite reliable results.

Figure 2$a$ gives the extrapolated values of $(M_1-2M)/\eel$ along
with the results obtained in \cite{ham1}.
\begin{figure}[p]
\begin{center}
\hskip -9mm \epsfig{file=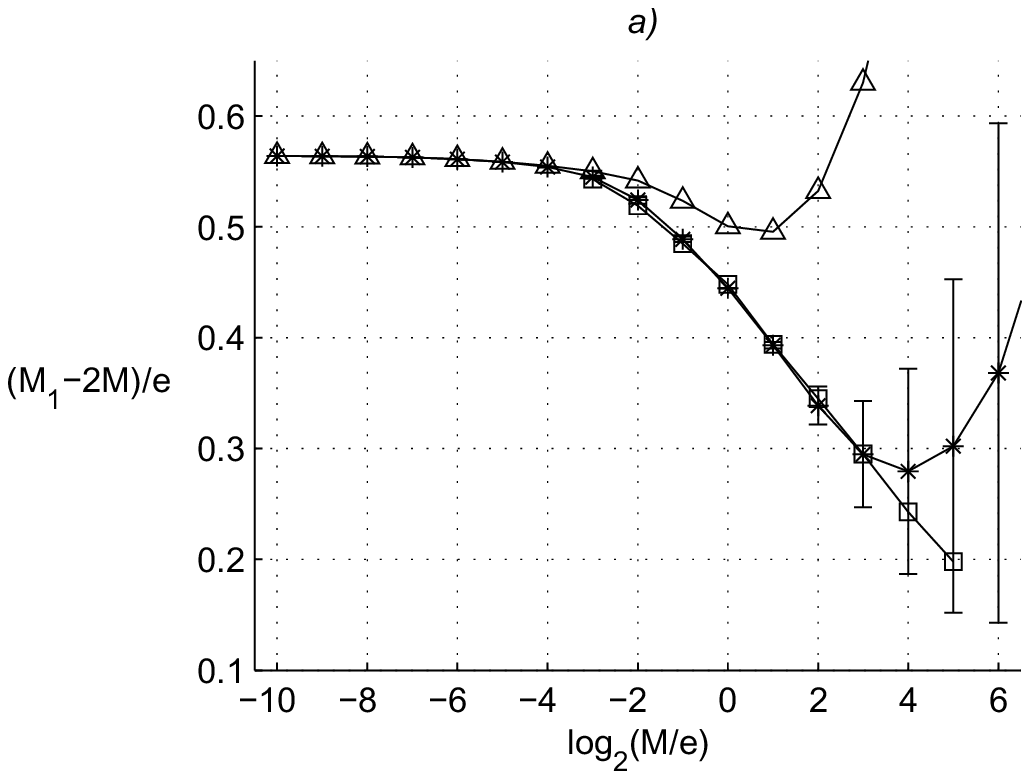,height=10.5cm}
\epsfig{file=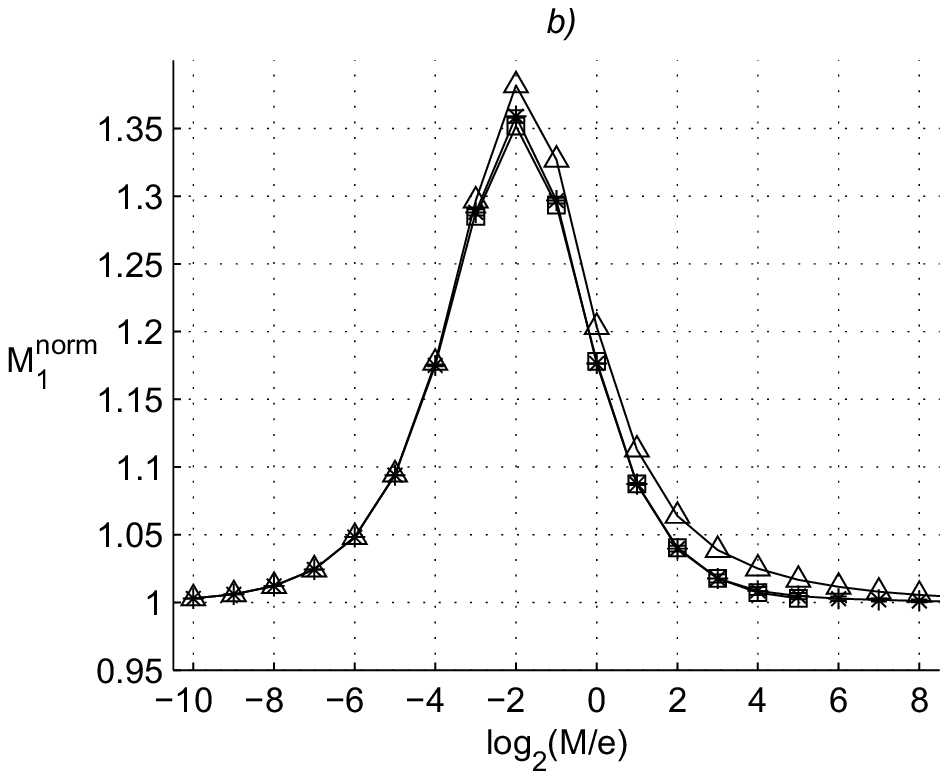,height=9.8cm}
\caption{
Calculated mass $M_1$ of the vector bound state at $\hat\te=\te=0$;
$*$ -- results obtained by extrapolation to the limit $N\to\infty$,
$\bigtriangleup$ -- results of the calculation at $N=30$,
$\Box$ -- results borrowed from \protect\cite{ham1}.
}
\end{center}
\end{figure}
The displayed errors were
found from the corresponding values of the relative error $\xi$. In
this figure, we also present the results corresponding to the
maximum accessible value of $N=30$ -- that is, the results
obtained without extrapolation. It can be seen that these results
are accurate only at small values of $M/\eel$; at the same time, the
extrapolated values give a very good result up to $M/\eel=8$. For
$M/\eel>8$, the extrapolated values reproduce a correct result
within the error, which begins growing fast in this region. This
is because the ratio $M/\eel$ becomes large in this region, with the
result that the absolute error of the difference $(M_1-2M)/\eel$
appears to be large even at a small relative error in the
calculated quantity $M_1^2/\eel^2$. In order to depict the calculated
results over the entire wide region of $M/\eel$, it is convenient to
plot the normalized values
 \disn{5.1}{
M^{\rm norm}_i=\frac{M_i}{\sqrt{m^2+(2M)^2}},
\nom}
as was proposed in \cite{adam2}. These normalized values possess
the property that $M^{\rm norm}_1\to 1$ both in the limit $M\to 0$ and
in the limit $\eel\to 0$. In Fig.~2$b$, we show the curves
for $M^{\rm norm}_1$ that correspond to those in Fig.~2$a$. One
can see that the extrapolated values agree with the results from
\cite{ham1} to a high precision over the entire range of $M/\eel$.

In Fig.~3, the extrapolated values of $M^{\rm norm}_1$ that were
calculated on the basis of the corrected Hamiltonian (\ref{3.1}) are
contrasted against the analogous values corresponding to the
Hamiltonian obtained upon the naive light-front canonical
quantization of the fermion theory specified by Eq.~(\ref{2}) that is,
to expression (\ref{3.1}) where one discards the second term
[see the
comment before formula (\ref{3.8})] and those corresponding to the
Hamiltonian obtained upon the naive light-front canonical
quantization of the boson theory specified by Eq.~(\ref{3.0}) that is, to
expression (\ref{3.1}) where one discards the third term
and replaces $\hat\te$
by $\te$, as was shown in \cite{shw2,tmf02}.
\begin{figure}[p]
\begin{center}
\epsfig{file=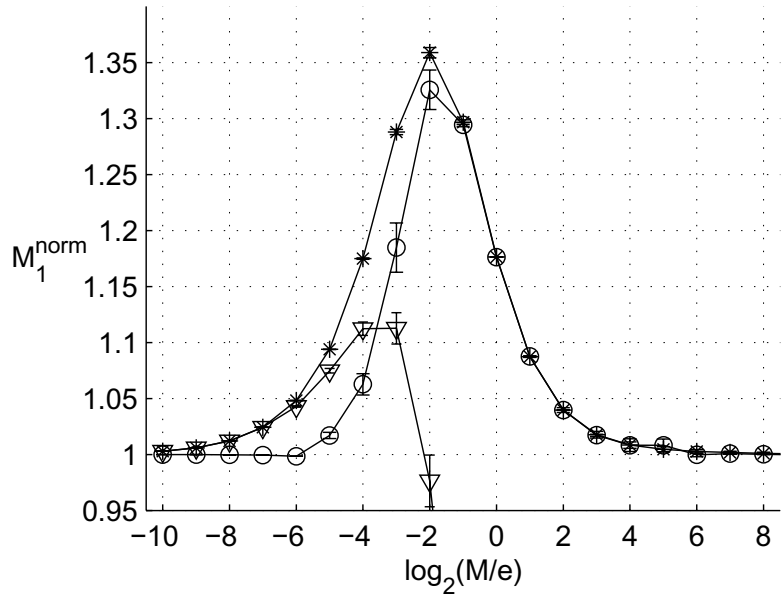,height=8.1cm}
\caption{
Calculated mass $M_1$ of the vector bound state at $\hat\te=\te=0$;
$*$ --  results obtained with the aid of the corrected
light-front Hamiltonian,
$\circ$ and $\bigtriangledown$ -- result
obtained by means of the naive light-front
quantization of, respectively, the fermion and the boson
formulation of the theory.
}
\end{center}
\end{figure}
One can see from this figure that
the naive light-front canonical quantization of the fermion
formulation of the theory gives good results at large values of
the ratio $M/\eel$ -- that is, in the region of weak coupling-while the
analogous quantization of the boson formulation of the theories
gives good results at small values of this ratio -- that is, in the
region of strong coupling. In order to obtain a light-front
Hamiltonian that would provide good results at any values of the
ratio $M/\eel$, it is necessary to implement the procedure of
correcting the naive Hamiltonian, as was done in \cite{shw2,tmf02}.

In Fig.~4, we give the curves that represent the mass $M_2$ of the
scalar bound state and which are analogous to those in Fig.~2$a$
for the mass $M_1$ of the vector bound state.
\begin{figure}[p]
\begin{center}
\hskip -12mm \epsfig{file=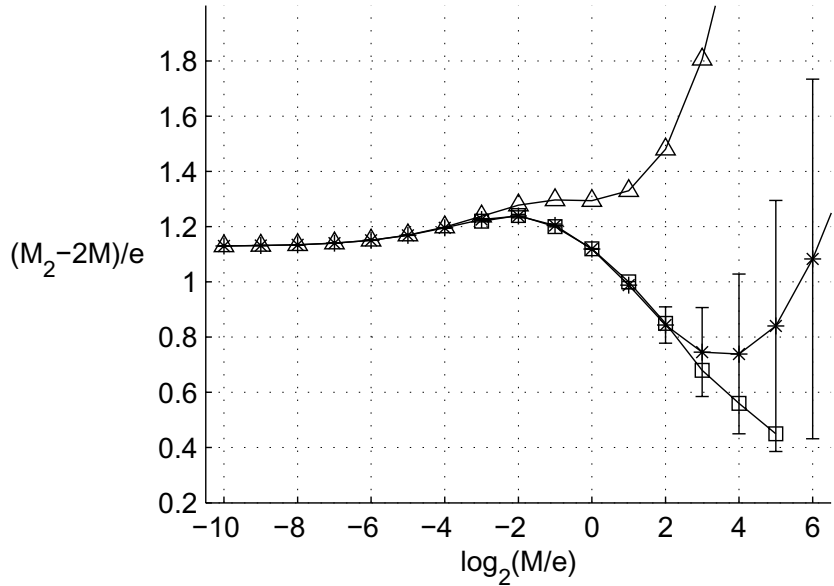,height=8.1cm}
\caption{
Calculated mass $M_2$ of the scalar bound state at $\hat\te=\te=0$;
$*$ -- results obtained by means of an extrapolation to the
limit $N\to\infty$, $\bigtriangleup$ -- results corresponding to $N=30$,
$\Box$ --  results borrowed from \protect\cite{ham1}.
}
\end{center}
\end{figure}
The behavior of the
curves is identical to that in the case of the vector state: the
values corresponding to $N=30$ give good results only at small
values of $M/\eel$, while the extrapolated values give very good
results up to $M/\eel=4$ and reproduce the correct result within the
error for $M/\eel>4$.

{\bf 4.2. Case of $\hat\te=\te=\pi$.}

The value of $\te=\pi$ is of particular importance in the theory.
It was predicted in \cite{24} that, at $\te=\pi$, a phase transition
occurs in the theory at some value of the ratio $M/\eel$, so-called
semi asymptotic fermions appearing in the region below the phase
transition. In the region above the phase transition, as well as
in the case where $\te\ne\pi$, confinement takes place. More recent
calculations, performed for $\te=\pi$ revealed (see, for example,
\cite{ham2}) that the phase transition occurs at $M/\eel=0.33$.

By means of lattice calculations, the mass of the lowest state in
the electron-positron (two-particle) sector as a function of $M/\eel$
at $\te=\pi$ was studied in \cite{ham2}. Within our approach, this mass
corresponds to the quantity $M_1$.

In Fig.~5, the extrapolated values of $M_1/\eel$ are given along with
the results reported in \cite{ham2}.
\begin{figure}[htbp]
\begin{center}
\epsfig{file=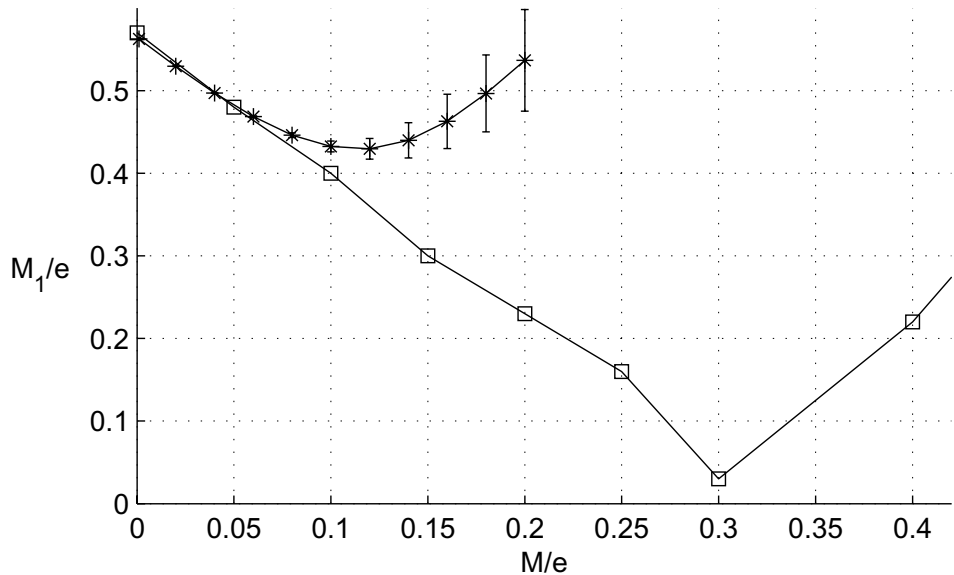,height=8cm}
\caption{
Calculated mass $M_1$ of the lowest bound state at $\hat\te=\te=\pi$;
$*$ -- results obtained by means of an extrapolation to the
limit $N\to\infty$, $\Box$ -- results borrowed from \protect\cite{ham2}.
}
\end{center}
\end{figure}
One can see that, at small values
of the ratio $M/\eel$, these results agree well, but that this
agreement deteriorates as the ratio $M/\eel$ grows. For $M/\eel>0.14$,
the extrapolation error $\xi$ [see formula (\ref{4.6})] exceeds 0.1;
therefore, the corresponding points on the graph are not quite
reliable. For $M/\eel>0.21$, there sharply appear very large
oscillations in the behavior of the corresponding functions
$P_1(n)$ describing the dependence of the extrapolated values on
the degree of the approximating polynomial [see Section~3, the
text before Eq.~(\ref{4.6})]. At $M/\eel=0.4$, the oscillations virtually
disappear, whereupon the dependence begins displaying a linearly
decreasing character. By way of example, this dependence at $M/\eel=0.5$
is depicted in Fig.~l$c$, while the corresponding dependence of
the quantity $M_1^2/\eel^2$ on $N$ (it is also manifestly linear) is shown in
Fig.~1$d$. For $M/\eel>0.2$, the extrapolation error $\xi$ exceeds 0.5.

The linear decrease in $M_1^2/\eel^2$ with increasing $N$ gives
sufficient grounds to assume that this quantity tends to
$-\infty$ in the limit $N\to\infty$. This means that, at the
above values of the ratio $M/\eel$ and the parameter $\hat\te$,
the spectrum of the Hamiltonian in (\ref{3.1}) upon the removal
of the regularization appears to be not bounded from below, so
that the theory specified by this Hamiltonian is incorrect. This
situation is possible in the case where there arise effects that
are purely nonperturbative from the point of view of perturbation
theory in the fermion mass $M$ since the Hamiltonian in (\ref{3.1}) was
constructed by analyzing such a perturbation theory (in all
orders). Obviously, the presence of the aforementioned phase
transition is a nonperturbative effect in this case. One can
conclude that above the phase-transition point ($M/\eel=0.33$), the
theory generated by the light-front Hamiltonian (\ref{3.1}) becomes
incorrect; at the same time, the original theory in Lorentz
coordinates, which is specified by Eq.~(\ref{2}), remains correct, this
being corroborated by the results reported in \cite{ham2} for the region
above the phase-transition point.

The appearance of the aforementioned strong oscillations of the
functions $P_1(n)$ in the range $0.2<M/\eel<0.4$ is likely to be
associated with the proximity of the phase-transition point,
where the regularization, which is parametrized by the number
$N$, can distort the theory more strongly than as usual. It can
also be conjectured that a sizable deviation of the extrapolated
values of $M_1/\eel$ from the results of paper \cite{ham2} in the
upper part of the region $M/\eel<0.2$ is due to the same factor.

{\bf 4.3. Case of Intermediate Values of $\hat\te$.}

As was indicated above, the quantity $\hat\te$ is a function of the ratio
$M/\eel$ and the parameter $\te$, this function being specified in the
form of an infinite series in $M$. Therefore, the relation between
$\hat\te$ and $\te$ is a priori unknown [this is not so only in the
particular cases of $\te=0,\pi$ (see above)]. In principle, this
relation can be sought by comparing the mass spectrum calculated
on the basis of the light-front Hamiltonian (\ref{3.1}), which depends on
$\hat\te$, and the spectrum calculated in Lorentz coordinates at a
specific vacuum angle $\te$.

It can be shown that the mass spectrum of the theory is invariant
under the reversal of the sign of the quantities $\hat\te$ and $\te$ (it
should be recalled that $\hat\te$ is an odd function $\hat\te$).
This can be seen
most straightforwardly in the boson form of the theory (\ref{3.0}),
where the reversal of the sign of $\te$ is equivalent to the
replacement ($\f$ by $-\f$, which does not change the mass spectrum.
However, this invariance can be directly seen from the
Hamiltonian in (\ref{3.1}). One can show that it does not change if we
reverse the sign of $\hat\te$ and simultaneously interchange the
operators $d_n$ and $b_{n+1}$ (in this case, $\om$ is replaced by $-\om$).
This is a unitary transformation and does not change the mass
spectrum. Thus, it is sufficient to perform calculations of the
mass spectrum for the case where the parameter $\hat\te$ lies between 0
and $\pi$.

For the lowest bound state, the mass $M_1$ calculated in this way
and normalized according to (\ref{5.1}) is displayed in Fig.~6.
\begin{figure}[htbp]
\begin{center}
\epsfig{file=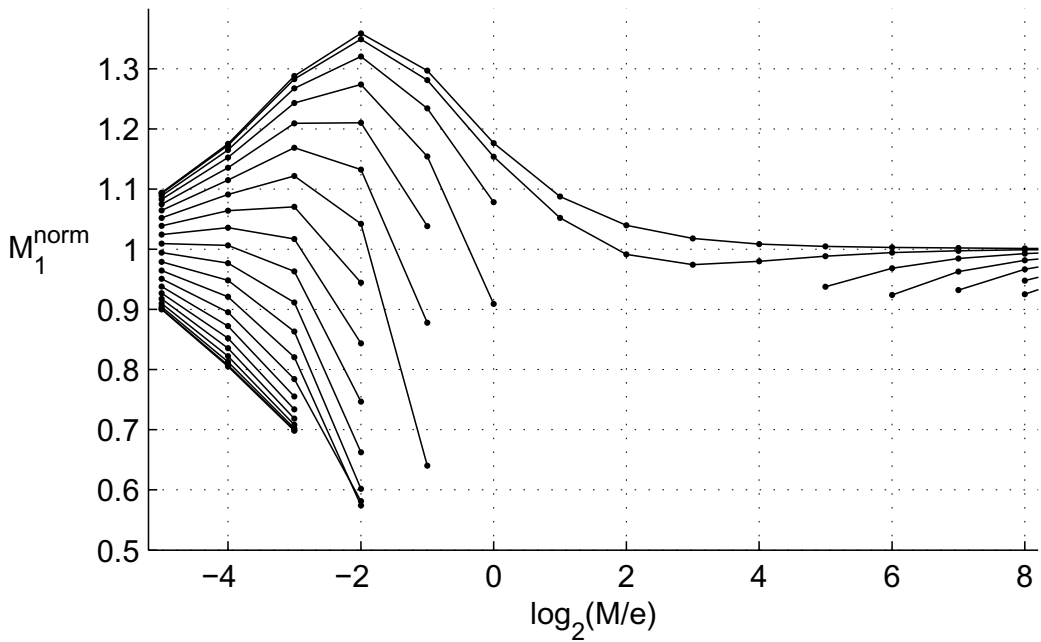,height=8.7cm}
\caption{
Calculated mass $M_1$ of the lowest bound state as a function of $M/\eel$
at various values of the parameter $\hat\te$:
$0,0.05\pi,0.1\pi,\dots,\pi$. As
the parameter $\hat\te$ is increased, the curve is shifted downward.
}
\end{center}
\end{figure}
Each curve corresponds to a fixed value of the quantity $\hat\te$ from the set
$0,0.05\pi,0.1\pi,\dots,\pi$. Each successive curve (for increasing
$\hat\te$) lies below the preceding one. In the cases of $\hat\te/\pi=0$ and
$\hat\te/\pi=0.05$ (the first and the second curve from above), the
extrapolation error $\xi$ [see (\ref{4.6})] does not exceed 0.1 at any
values of $M/\eel$.

In the case of $\hat\te/\pi=0.1$, $\xi>0.1$ at $M/\eel$ values in the range
between 2 and $2^4$; therefore, it is meaningless to plot the
corresponding points on the graph, so that the curves decompose
into two parts. In this region, the corresponding functions $P_1(n)$
(used in the extrapolation) display a manifest linearly
decreasing character. This situation is similar to that
considered in Subsection~4.2 for the case of $\te=\pi$. It can be
concluded that, in the region being considered, the theory
specified by the Hamiltonian in (\ref{3.1}) becomes incorrect upon the
removal of the regularization, and it can be assumed that, in the
vicinity of the point $M/\eel=2$, there exists some nonperturbative
effect as in the case of $\te=\pi$ (see Subsection~4.2).

In principle, it can be assumed that a nonperturbative effect
exists in the vicinity of the point $M/\eel=2^4$ as well, above which
the error $\xi$ again falls below 0.1. However, it seems more
probable that, in the region of large values of $M/\eel$, the
Hamiltonian remains unbounded from below upon the removal of the
regularization, but the decrease in the mass of the lowest state
with increasing regularization parameter $N$ is so slow that $\xi$
appears to be less than 0.1. This is favored by the dependence
$P_1(n)$ which, at large values of $M/\eel$, has the form of a linear
function with a moderate slope that yields $\xi<0.1$. As was
discussed at the end of Section~3, there is unfortunately no
method for discriminating between the situations where, in the
limit $N\to\infty$, there exists the limit of the mass of the lowest
state and where this mass tends to $-\infty$.

In the cases of $\hat\te/\pi=0.15,\dots,0.3$, the situation is perfectly
analogous to that in the case of $\hat\te/\pi=0.1$ (only the width of
the region where $\xi>0.1$ changes), which was considered
immediately above, while, at $\hat\te/\pi=0.35,\dots,1$ the difference
consists in that the region where $\xi<0.1$ at large values of $M/\eel$
is not reached at the $M/\eel$ values considered here.

In Fig. 7, the points where the extrapolation error $\xi$ is less
than 0.1 are shown in the plane spanned by the parameters $M/\eel$ and
$\hat\te$.
\begin{figure}[htbp]
\begin{center}
\epsfig{file=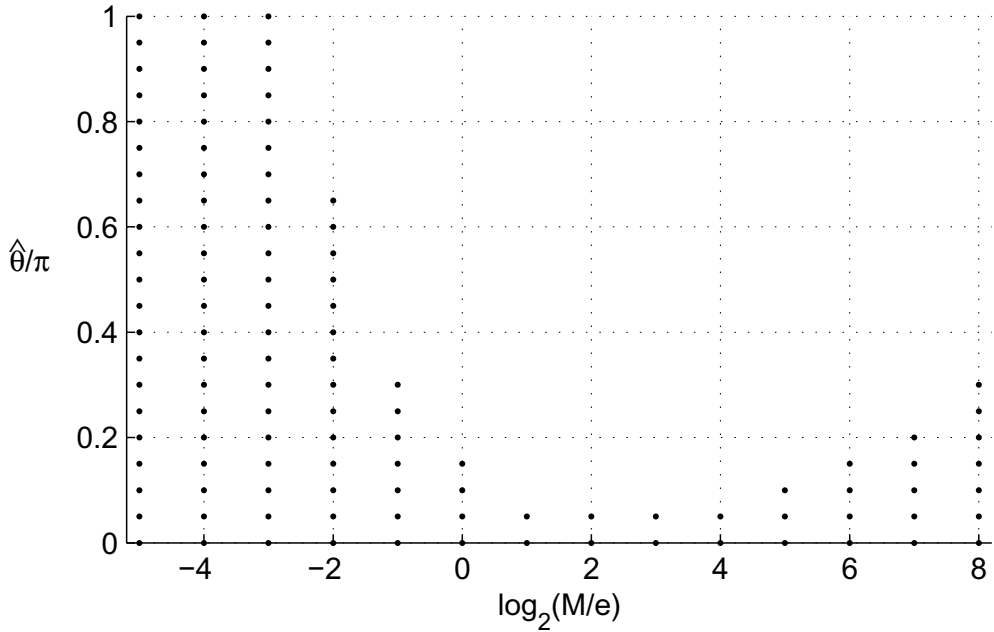,height=8.7cm}
\caption{
Set of pairs of $M/\eel$ and $\hat\te$ values at which
the extrapolation error $\xi$ is less than 0.1.
}
\end{center}
\end{figure}
It is natural to assume that, in the region where there are no
points in the figure, the light-front Hamiltonian (\ref{3.1}) is
unbounded from below upon the removal of the regularization, $N\to\infty$.

\section{Conclusion}
A numerical nonperturbative calculation of the mass spectrum of
QED-2 (massive Schwinger model) has been performed by the method
of discrete light-cone quantization. In doing this, use has been
made of the corrected light-front Hamiltonian (that is, that
which generates a theory that is equivalent in all orders of
perturbation theory in the fermion mass $M$ to the
Lorentz-covariant formulation of QED-2) constructed in \cite{shw2,tmf02}.
The results obtained in this way have been compared with the
results of the numerical calculations on a lattice in Lorentz
coordinates from \cite{ham1,ham2}.

Since actual calculations are performed at a finite value of the
infrared-regularization parameter $N$, a method has been proposed
for extrapolating the results of these calculations to the region
of $N$ tending to infinity, this corresponding to the removal of
this regularization. As the result of this extrapolation, it
becomes clear that, at some values of the parameters of the
theory, its mass spectrum is not bounded from below in the limit
where the regularization is removed. This occurs only at rather
large values of the fermion mass $M$; therefore, it is natural to
assume that, in this region, the Hamiltonian used becomes
incorrect, since it was constructed via an analysis of
perturbation theory in $M$.

The calculations have been performed over a wide range of the
fermion mass $M$ for all values of the Hamiltonian parameter $\hat\te$,
which is a function of the ratio $M/\eel$ and the vacuum angle $\te$.
This function, which is a priori unknown possesses the property that
it vanishes at $\te=0$ and is equal to $\pi$ at $\te=\pi$.

At zero value of the vacuum angle $\te$, the resulting spectrum is
bounded from below at any value of $M$, and the values obtained in
this case for the two lowest bound states reproduce well the
results reported in \cite{ham1}.

At the vacuum-angle value of $\te=\pi$, which is special for the
theory being considered, the masses found here for the lowest
bound state agree well with the results of paper \cite{ham2} for
rather low values of $M$; as the mass $M$ increases, there first
arises a discrepancy, whereupon the spectrum of the theory become
unbounded from below. Since the $M$ value at which this takes place
is approximately equal to that at which there occurs a phase
transition in the theory (see, for example, \cite{ham2})), it would be
reasonable to assume that, at $M$ values above the phase-transition
point, the light-front Hamiltonian used here, which was
constructed on the basis of an analysis of perturbation theory in
$M$, becomes incorrect.

Our calculations have revealed that, in the case of QED-2, the
procedure employed in \cite{shw2,tmf02} to construct the
corrected light-front Hamiltonian leads to a Hamiltonian that
yields good results in nonperturbative calculations. By exploring
the question of how the spectrum of the Hamiltonian changes in
response to changes in the parameters of the theory from the
perturbative region of their values, one can determine the
boundaries of the applicability region of this Hamiltonian -- that
is, to find the region where it is necessary to take additionally
into account nonperturbative (for example, vacuum) effects. This
may be of use in studying more realistic gauge field theories.

The LF Hamiltonian (\ref{3.1}), used in the present paper
for the calculation of bound state mass, includes the operator
$\om$, which has no simple expression in terms of field operators.
It is defined only by its properties (\ref{3.5})-(\ref{3.7}).
Due to this fact the expression for the Hamiltonian depends essentially
on the form of the regularization, i.e. $|x^-|\le L$ and
antiperiodic boundary conditions in $x^-$ for the
field $\psi$.
Now we have found a possibility to rewrite the expression (\ref{3.1})
in such a way that it contains only fermion field operators,
and describes in the limit of removing the regularizations the same theory
as the Hamiltonian (\ref{3.1}).
This new expression has at $\hat\te=\te=0$ the following form:
 \disn{new1}{
H=\int\limits_{-L}^Ldx^-\ls\frac{e^2}{2}\ls \dd_-^{-1}
[\psi^+\psi]\rs^2
+\frac{eMe^C}{4\pi^{3/2}}\ls d_0^+d_0+b_1^+b_1\rs-
\frac{iM^2}{2}\psi^+\dd_-^{-1}\psi\rs.
\nom}
Preliminary calculations of the mass spectrum, produced by this Hamiltonian,
show that in the limit of removing the regularization, $L\to\infty$, results
indeed coinside, with a good accuracy, with those for the bound state mass
spectrum found here for the Hamiltonian (\ref{3.1}).
Work on studing of the Hamiltonian (\ref{new1}) spectrum and also
on the constructing the analogous expression for the Hamiltonian
at $\hat\te\ne0$  will be continued in future.

\vskip 1em
{\bf Acknowledgments.}
The work was supported by the administration of
St.~Petersburg and the Ministry of Education of the Russian
Federation, grant no.~PD03-1.2-35 (S.A.P.),
and by the Russian Foundation for Basic Research,
grant no.~05-02-17477 (S.A.P. and E.V.P.).

\setcounter{equation}{0}
\renewcommand{\theequation}{{\rm A}.\arabic{equation}}
\vskip 1em
\hfill {\bf Appendix}
\vskip 1em

Let us find out how the parameter $\hat\te$, which is a function of the
ratio $M/\eel$ and the vacuum angle $\te$, appears in the light-front
Hamiltonian (\ref{3.1}). It was shown in \cite{shw2,tmf02} that the coefficient of
the operator $e^{i\om}d_0^+$ in the integrand on the right-hand
side of (\ref{3.1})
is the limit of the quantity $-B^*$ ($*$ denotes the operation of
complex conjugation) in the limit $w\to\infty$ (which corresponds to
removing the intermediate ultraviolet regularization), where $B$ is
given by
 \disn{p1}{
B=-\frac{1}{2w}+\sqrt{\frac{1}{4w^2}+\frac{A'}{w}-A''^2}+iA''.
 \nom}
Here, $A'$ and $A''$ are, respectively, the real and the
imaginary part of the sum (calculated in Lorentz
coordinates and in all orders in $\g$, including the first order)
 \disn{p2}{
A=\frac{\g}{2}e^{i\te}+\sum_{k=2}^\infty A_k\g^k
 \nom}
of all connected diagrams of the boson theory specified by the
Lagrangian (\ref{3.0}) which have a property that all their
external lines connect with the one vertex $\frac{\g}{2}e^{i\te}$
[the propagators of the
external lines and a common factor that depends on their number
are not included in the definition of $A$, as can be seen from the
first term in (\ref{p2})]. In \cite{shw2}, it was established that, in the
limit $w\to\infty$, the quantities $A_k$ for $k\ne 2$ are finite, while
$A_2$ behaves as
 \disn{p3}{
 A_2=\frac{\g^2}{4}w+const.
 \nom}

Expression (\ref{p1}) was deduced from an analysis of
perturbation theory in $\g$ (in all orders). As a matter of fact,
the series that can be obtained by substituting expansion
(\ref{p2}) into (\ref{p1}) provides the definition of the
quantity $B$. One can see that the first term of this series is
linear in $\g$ and, at large $w$, its radius of convergence
varies in proportion to $1/w$. As was mentioned above, it is
necessary to find the limit of the quantity $B$ for $w\to\infty$.
Since the aforementioned radius of convergence tends to zero in
this limit, it is obvious that, before going to the limit, the
quantity $B$ must be continued analytically in $\g$ to the region of
positive $\g$ values, which lie beyond the disk determined by this
radius of convergence. For this, we determine the behavior of
radicand on the right-hand side of (\ref{p1}) at large $w$ and $\g$ of
about $1/w$. Taking into account (\ref{p3}), we obtain
 \disn{p4}{
 \sqrt{\frac{1}{4w^2}+\frac{A'}{w}-A''^2}=\frac{1}{2w}
 \sqrt{(1+w\g\cos\te)^2+O\ls\frac{1}{w}\rs}.
 \nom}
From here, we find that there exist two branch points, whose
positions are given by the formula
 \disn{p5}{
 \g_{1,2}=-\frac{1}{w\cos\te}+O\ls \frac{1}{w^{3/2}}\rs\! .
 \nom}
It can be concluded from formulas (\ref{p4}) and (\ref{p5}) that, in the
limit $w\to\infty$, the sought analytic continuation of the quantity $B$
has the form
 \disn{p6}{
B=\si(\cos\te)\sqrt{\frac{\g^2}{4}-A''^2}+iA''=
\frac{\g}{2}e^{i\hat\te}.
 \nom}
The form of this expression corresponds to the form of the
coefficient of the operator $e^{i\om}d_0^+$ in Hamiltonian (\ref{3.1}).
Expression
(\ref{p6}) differs by the presence of the sign function $\si(\cos\te)$ from
the corresponding expression presented in \cite{shw2,tmf02}, where the
features of the analytic continuation of $B$ were not taken
into account. In the first order in $\g$ (and, hence, in $M$), we find
from (\ref{p6}), with the aid of expansion (\ref{p2}), that $\hat\te=\te$.

Upon the Euclidean rotation in the diagrams determining the
quantity $A$, it becomes clear $A$ is a real-valued function of $m$, $\g$,
$e^{i\te}$ and $e^{-i\te}$. It follows from here that the reversal of the
sign of the parameter $\te$ is equivalent to the complex conjugation
of the quantity $A$, this quantity being real-valued at $\te=\pi$. By
using these facts and formula (\ref{p6}), one can easily see
that $\hat\te$ is
an odd function of $\te$ and, in addition, that $\hat\te=\pi$ at $\te=\pi$.

\end{document}